\documentclass[10pt,letterpaper]{article}
\usepackage{opex3,amsmath,bm}
\usepackage[colorlinks=true,allcolors=blue]{hyperref}

\newcommand{\vect}[1]{\ensuremath{\bm{\mathrm{#1}}}}
\newcommand{\sinc}[1]{\ensuremath{\mathrm{sinc}\, #1}}
\newcommand{\ket}[1]{|#1\rangle}

\begin{document}
\bibliographystyle{unsrt}

\title{Cancellation of atmospheric turbulence effects in entangled two-photon beams}%

\author{M. V. da Cunha Pereira, L. A. P. Filpi and C. H. Monken$^*$}%
\address{Departamento de F\'isica, Universidade Federal de Minas Gerais, Caixa Postal 702, \\ Belo Horizonte, MG 30123-970, Brazil}
\address{$^*$Corresponding author: \href{mailto:monken@fisica.ufmg.br}{monken@fisica.ufmg.br}}%
\date{\today}
\begin{abstract}
Turbulent airflow in the atmosphere and the resulting random fluctuations in its refractive index have long been known as a major cause of image deterioration in astronomical imaging and figures among the obstacles for reliable optical communication when information is encoded in the spatial profile of a laser beam. Here we show that using correlation imaging and a suitably prepared source of photon pairs, the most severe of the disturbances inflicted on the beam by turbulence can be cancelled out. Other than a two-photon light source, only linear passive optical elements are needed and, as opposed to adaptive optics techniques, our scheme does not rely on active wavefront correction.
\end{abstract}
\ocis{(010.1300) Atmospheric propagation; (270.5565) Quantum communications}



\section{Introduction}
Fluctuations of the atmospheric refractive index due to turbulent air flow have long been known as a major cause of image deterioration in astronomical imaging. Such fluctuations figure among the main obstacles to reliable optical communication \cite{Andrews2005}, by causing signal attenuation and distortion. One possible way to avoid turbulence effects in optical communications is to encode information in the polarization degree of freedom, since it is well known that polarization is marginally affected by clear air propagation disturbances \cite{Fante1975,Fedrizzi2009}. However, due to the two-dimensional nature of polarization states, the amount of information carried by a single pulse is limited to one bit, or to one qubit per photon in the quantum case. On the other hand, transverse spatial degrees of freedom of higher-order beams allow for spaces of much larger dimensionality to be accessed \cite{Molina-Terriza2001,Gibson2004,Neves2005,Pors2008,Walborn2010,Wang2012}, and these have been gaining prominence even in optical fiber communications, through the use of few-mode fibers and spatial division multiplexing for increased communication capacity\cite{Randel2011}. In the case of free-space (not vacuum) propagation however, there is the problem that information encoded in photonic spatial degrees of freedom are typically severely degraded by atmospheric turbulence \cite{Paterson2005,Malik2012}. 

Entanglement has been exploited in numerous situations to cancel unwanted effects such as group-velocity dispersion \cite{Nasr2003}, spatial aberrations and dispersion \cite{Bonato2008,Simon2009,Chan2011,Dixon2011} and, with the use of a phase-conjugating mirror, a proposal has been made even to mitigate atmospheric turbulence phase distortions \cite{Simon2011}. It has also been recently shown in ref. \cite{Prevedel2011} that the strong correlations observed in entangled states can assist in performing classical communication under the effect of specific sources of error. We show here a new scheme that does not resort to active compensation but, with the aid of the correlations present in a spontaneous parametric down-conversion (SPDC) source, makes spatial information immune to all odd-order spatial aberrations (both in amplitude and phase) caused by refractive index fluctuations in the atmosphere. Among these is the wavefront tilt, which has long been known \cite{Fried1965} for being the most deleterious in terms of loss in resolution when long-exposure imaging is used \cite{Fried1966}, accounting for as much as $80\% $ of the total wavefront distortion \cite{Andrews2005}.

\section{Theory}
Our basic scenario (see fig \ref{fig:setup}) is a two-photon light beam of wavenumber $k$ generated by collinear type I SPDC propagating horizontally in the $z$ direction, impinging on a receiver (detector) of aperture $A$ located at a distance $L$ from the transmitter (the nonlinear crystal). Throughout the propagation path, the field undergoes the effects of atmospheric turbulence (refractive index fluctuations). The SPDC pump beam is a laser beam of wavenumber $k_p=2k$, focused on $A$. The two-photon state generated by collinear type I SPDC, neglecting birefringence effects in the nonlinear crystal, in the monochromatic and paraxial approximations is proportional to \cite{Walborn2010}
\begin{equation}
\label{phi}
\int\!d^2q_1\int\!d^2q_2\, \mathcal{E}_p(\vect{q}_1+\vect{q}_2)\,\sinc{\beta|\vect{q}_1-\vect{q}_2|^2} \ket{\vect{q}_1}\ket{\vect{q}_2},
\end{equation}
where $\vect{q}_1$ and $\vect{q}_2$ are the transverse ($xy$) components of the down-converted wave vectors, $\mathcal{E}_p$ is the pump beam plane wave spectrum, $\beta=\tau/4k_p$, $\tau$ is the nonlinear crystal thickness, and $\ket{\vect{q}_1}$ and $\ket{\vect{q}_2}$ are one-photon states in plane wave modes defined by $\vect{q}_1$ and $\vect{q}_2$. This state gives rise to the following two-photon detection amplitude at the point $\vect{r}_1=\vect{r}_2=\vect{r}=(\vect{\rho},L\hat{\vect{z}})$:
\begin{equation}
\label{ampli}
\mathcal{A}^{(2)}(\vect{r},\vect{r})\propto\int\!d^2\rho^\prime_1\int\!d^2\rho^\prime_2\, 
E_p\left(\frac{\vect{\rho}^\prime_1+\vect{\rho}^\prime_2}{2}\right)V(\vect{\rho}^\prime_1-\vect{\rho}^\prime_2) \exp{\left[\frac{ik}{2L}\left(|\vect{\rho}^\prime_1-\vect{\rho}|^2+|\vect{\rho}^\prime_2-\vect{\rho}|^2\right)\right]},
\end{equation}
where $E_p$ is the Fourier transform of $\mathcal{E}_p$, that is, pump beam profile on the $z=0$ plane (nonlinear crystal) and $V$ is the Fourier transform of the sinc function. 

In the presence of turbulence, we have to include complex random phase factors $\psi$ for each down-converted photon in \eqref{ampli}, which leads to 
\begin{eqnarray}
\label{ampliT}
\mathcal{A}_T^{(2)}(\vect{r},\vect{r})&\propto&\int\!d^2\rho^\prime_1\int\!d^2\rho^\prime_2\, E_p\left(\frac{\vect{\rho}^\prime_1+\vect{\rho}^\prime_2}{2}\right)V(\vect{\rho}^\prime_1-\vect{\rho}^\prime_2) \exp{\left[\frac{ik}{2L}\left(|\vect{\rho}^\prime_1-\vect{\rho}|^2+|\vect{\rho}^\prime_2-\vect{\rho}|^2\right)\right]}\nonumber\\
&&\times \exp\left[\psi(\vect{\rho}^\prime_1,\vect{\rho};k)+\psi(\vect{\rho}^\prime_2,\vect{\rho};k)\right],
\end{eqnarray}
where each complex phase factor $\psi(\vect{\rho}^\prime,\vect{\rho};k)$ represents a random distortion, both in phase and amplitude, of a spherical wave with wavenumber $k$ originated in $\vect{\rho}^\prime$ on the source plane, and observed in $\vect{\rho}$ on the observation plane ($z=L$). If we consider that the nonlinear crystal is thin enough (typically, a few millimeters), the $\mathrm{sinc}$ function in Eq. \eqref{phi} is broad in  $\vect{q}_1-\vect{q}_2$ and its Fourier transform $V$ can be approximated by a delta function $\delta(\vect{\rho}^\prime_1-\vect{\rho}^\prime_2)$. Then, $\mathcal{A}_T^{(2)}$ can be approximated by
\begin{equation}
\label{ampli2}
\mathcal{A}_T^{(2)}(\vect{r},\vect{r})\propto\int\!d^2\rho^\prime\, E_p\left(\vect{\rho}^\prime\right) \exp{\left[\frac{ik_p}{2L}|\vect{\rho}^\prime-\vect{\rho}|^2\right]}
\exp\left[\psi(\vect{\rho}^\prime,\vect{\rho};k_p)\right],
\end{equation}
where in the passage from  \eqref{ampliT} to \eqref{ampli2} we used the fact that $k_p=2k$ and $2\psi(\vect{\rho}^\prime,\vect{\rho};k)=\psi(\vect{\rho}^\prime,\vect{\rho};k_p)$, neglecting dispersion. From Eq. \eqref{ampli2} it is clear that the two-photon beam under turbulence behaves like the pump beam would in the same conditions.

We now show that the effects of turbulence can be mitigated if we make a coordinate inversion in one of the photons: $\vect{q}_2\rightarrow -\vect{q}_2$ and calculate the two-photon detection amplitude at the points $\vect{r}_1=\vect{r}=(\vect{\rho},L\hat{\vect{z}})$ and $\vect{r}_2=\vect{\tilde{r}}=(-\vect{\rho},L\hat{\vect{z}})$. All the steps above can be repeated, leading to
\begin{equation}
\label{ampli3}
\mathcal{A}_T^{(2)}(\vect{r},\vect{\tilde{r}})\propto\int\!d^2\rho^\prime\, E_p\left(\vect{\rho}^\prime\right) \exp{\left[\frac{ik_p}{2L}|\vect{\rho}^\prime-\vect{\rho}|^2\right]}
\exp\left[\psi(\vect{\rho}^\prime,\vect{\rho};k)+\psi(-\vect{\rho}^\prime,-\vect{\rho};k)\right].
\end{equation}

One can see from Eq. \eqref{ampli3} that the total perturbation is now symmetric with respect to the \textit{z} axis, that is to say, the antisymmetric part of $\psi$ is cancelled out. In particular, wavefront tilt, modeled by
$\psi_t(\vect{\rho}^{\prime},\vect{\rho};k)=(ik/2L)(\vect{\rho}^{\prime}-\vect{\rho})\cdot \vect{d}$,
where $\vect{d}$ is a random displacement, can be seen to vanish in Eq. \eqref{ampli3}. It is imperative for the cancellation to happen that the correlated photons propagate near-collinearly and close enough so that the random process describing turbulence $\psi$ is the same for both.

\section{Experiment}
\begin{figure}
\centering
\includegraphics{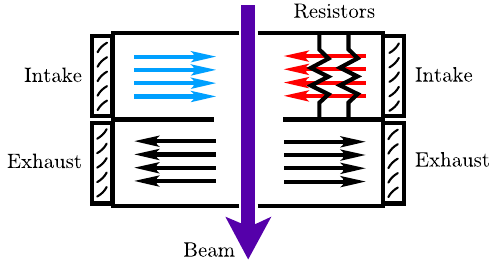}
\caption{(color online).  A turbulence chamber that emulates the turbulent mixing of air at different temperatures in the atmosphere.}
\label{fig:turbulador}
\end{figure}

To emulate the conditions present in the real atmosphere we adapted a tabletop turbulence generator put forward by Keskin \cite{Keskin2006}, depicted in figure \ref{fig:turbulador}. Cold and hot air fluxes are mixed inside an aluminum box, resulting in a random temperature field. Consequently, the refractive index inside the box fluctuates in both space and time \cite{Strohbehn1978}. Air is blown into the chamber by two fans, one of them having a resistor bank ($6.4 \; \Omega$) in front of it, which is driven by an adjustable power supply to dissipate up to 200 W.  Two other fans work as exhausts, to ensure that the process is stationary.
Each dissipated power corresponds to a different turbulence strength, which we characterize as follows: we expand and then focus a He-Ne laser beam with a lens system onto a charge-coupled device (CCD) camera, placed at a distance of 93 cm from the output lens. The beam has a wavenumber $k_l=2\pi/0.633=9.93$ $\mu$m$^{-1}$ and waist ($e^{-2}$ radius on the detection plane) $w_0= 58.3\ \mu$m. The turbulence chamber, positioned halfway between the beam expander and the CCD, causes the beam spot to undergo a random motion and distortion. All long-term average (over 55 s) transverse profiles remain Gaussian in shape, yet with a radius $w_{LT}$, which increases according to \cite{Andrews2005}
\begin{equation}
\label{wlt}
w^2_{LT}=w^2_0 \left(1+ \frac{k_l^{1/3}}{w_0^{5/3}} \gamma^2\right).
\end{equation}

Turbulence strength can be thus characterized by $\gamma = [7.75 L^{5/3} \int_0^L C_n^2(z) \left(1-z/L\right)^{5/3} dz]^{1/2}$, where $C_n^2$ is the refractive index structure constant. The parameter $\gamma$  is closely related to the more commonly used Rytov variance $\sigma_R^2= 1.23 C_n^2k_l^{7/6}L^{11/6}=0.423k_l^{7/6}L^{-5/6}\gamma^2$ when $C_n^2$ is constant through the propagation path. Kolmogorov spectrum for the refractive index spatial fluctuations $\Phi_n(\kappa)=0.033C_n^2\kappa^{-11/3}$ is assumed in the derivation of Eq. \eqref{wlt} \cite{Andrews2005}. Equation \eqref{wlt} allows us to determine $\gamma$ for each measured dissipated power as shown in Table~I.
\begin{table}[hb]
\centering
\begin{tabular}{c|c|c}
\hline\hline
Power (W) & $w_{LT}$($\mu$m) & $\gamma$($\mu$m)\\
\hline\hline
0   & 58.3 & 0\\
16.0  & 59.6 & 4.29 \\
34.5  & 62.7 &  7.99 \\
60.0  & 68.0 &  12.1 \\
95.0 & 75.4 & 16.6 \\
135.0 & 84.7 &  21.3 \\
185.5 & 99.5 &  27.9 \\
\hline
\end{tabular}
\caption{Calibration of the turbulence chamber. The parameter $\gamma$ is listed as a function of the power dissipated by the resistors.}
\end{table}

\begin{figure}[ht]
\centering
\includegraphics{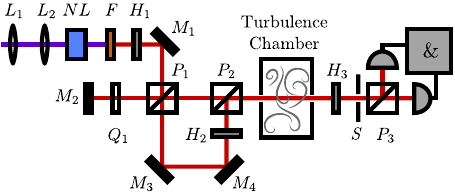}
\caption{(color online). \textbf{Experimental setup.} The two-photon source consists of the lens set $L_1$, $L_2$, the nonlinear crystal $NL$ and the uv filter $F$. The arrangement of optical elements from $H_1$ to $H_3$ is responsible for the control of coordinate inversion of one photon (see text). The detection system consists of an aperture $S$, a polarizing beam-splitter $P_3$ and two photon counters.}
\label{fig:setup}
\end{figure}

The experimental verification of expressions \eqref{ampli2} and \eqref{ampli3} was made in two steps, which we refer to as first experiment and second experiment, respectively. The setup is depicted in Fig. \ref{fig:setup}. A two-photon beam with wavelength of 650 nm is produced by degenerate spontaneous parametric down-conversion in a 5 mm long BiB$_3$O$_6$ (BiBO) nonlinear crystal (NL), such that the two-photon (coincidence) detection profile on the detection aperture $S$ is a Gaussian whose width is 60 $\mu$m, whereas the single-photon (intensity) profile width is of the order of 10 mm. The detection aperture is a 50 $\mu$m-wide slit oriented vertically. We note that refractive index fluctuations in air have spatial scales exceeding a millimeter \cite{Fante1975}, so that for $\rho < 50 \mu m$, the complex phase factors in Eq. \eqref{ampli3} satisfy $\psi(\vect{\rho}^{\prime},\vect{\rho};k) \approx \psi(\vect{\rho}^{\prime},\vect{0} ; k)$, and cancellation still effectively follows as if we were using point detectors centered in $\vect{\rho}=\vect{0}$. The detectors are avalanche photodiodes operating in photon counting mode with a coincidence resolving time of 5 ns. Interference band-pass filters centered at 650 nm with 10 nm bandwidth and coupling microscope objectives are used in front of each detector.

In the first experiment, all photons are made to propagate through the same path of an interferometer-like arrangement, so that the coincidence rate is described by Eq. \eqref{ampli2}, while in the second experiment all coincidence events correspond to each photon of a pair going through different arms. The arrangement is made such that there is an additional reflection in one of the arms, resulting in the momentum inversion necessary to obtain the conditions leading to Eq. \eqref{ampli3}. To switch between these configurations, the polarization of the photon pairs can be rotated before or after the turbulence to 45 degrees with respect to the horizontal plane, by means of two half-wave plates (HWP) $H_1$ and $H_3$. 

When $H_1$ is set to $0^\circ$, all photon pairs are reflected by the polarizing beam-splitter (PBS) $P_1$, back reflected with horizontal polarization by the combination of a quarter-wave plate $Q_1$ and a mirror $M_2$, and transmitted by PBS $P_2$. After crossing the turbulence chamber,   the photon pairs have their polarization rotated to $45^\circ$ by HWP $H_3$ and are split by PBS $P_3$ with probability 1/2. The split events are registered as coincidence counts. This situation corresponds to the first experiment.

For the second experiment, HWP $H_1$ rotates the photon polarizations to $45^\circ$ so that the photon pairs are split by $P_1$ with probability 1/2. The HWP $H_2$ in the bottom arm rotates the polarization from horizontal to vertical, so that all photons leave $P_2$ going towards the turbulence chamber. Because $H_3$ is now set to $0^\circ$, only the cases where the photon pairs are split by $P_1$ give rise to coincidence counts. Since a photon going through the bottom arm undergoes an additional horizontal reflection, a coordinate inversion is effected on the horizontal plane. 

\begin{figure}[ht]
\centering
\includegraphics{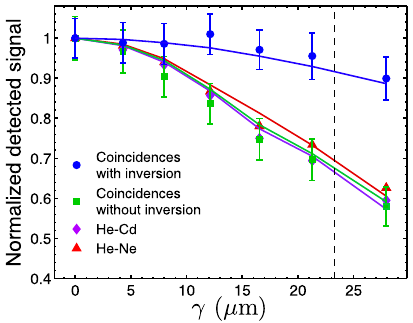}
\caption{(color online). \textbf{Experimental results.} Normalized detection counts plotted against turbulence strength as measured by $\gamma$. Plot markers and solid lines represent measured and simulated values, respectively. Blue circles: Normalized coincidence counts when the $x$ coordinate of one photon is inverted. Green squares: normalized coincidence counts with no coordinate transformation. Purple diamonds: integrated gray levels of the He-Cd laser long-term average intensity profiles in a 50 $\mu$m wide region of the CCD camera. Red triangles: Values used for the calibration of $\gamma$. Same procedure as for the He-Cd, but with the He-Ne laser.}
\label{fig:results}
\end{figure}

Distortions and beam deflections make the averaged coincidence transverse profile larger at the aperture plane and, because of the small width of the slit, the signal decreases. Results are shown in Fig. \ref{fig:results}, with normalized detection counts plotted against the turbulence strength, as measured by the parameter $\gamma$. Absolute coincidence rates without turbulence ($\gamma=0$) are 75 and 78 counts per second in the first and second experiments, respectively. We also included the measurements done with the He-Ne laser used to calibrate $\gamma$ and with the pump He-Cd laser. In this  latter measurement,  the laser beam was focused down to a waist radius of 58 $\mu$m onto a CCD, where the slit was previously positioned, as it was done with the He-Ne laser. For both the He-Ne and He-Cd measurements, we mimicked the $50\ \mu$m slit by integrating the gray level value over a region of interest defined by a vertical stripe 11 pixels wide in the CCD camera (which amounts to 51 $\mu$m). In both coincidence data sets (blue circles and green squares), each point is the result of an average over 100 samples of coincidence counts with a sampling time of 5 s, while for the He-Ne and He-Cd measurements (red triangles and purple diamonds) we recorded 11 sample videos of 5 s each. The error bars are the standard deviations over the samples. One can immediately see that the coincidences without inversion and pump beams perform rather similarly under turbulence, as expected from Eq. \eqref{ampli2}.

The plot is separated by a vertical dashed line into two distinct regions. The left and right parts correspond to the weak and strong turbulence regimes for the He-Ne laser, respectively. The separating line  is defined by the value of $\gamma$ such that $k_l^{1/3}w_0^{-5/3}\times \gamma^2 \approx 1.33$, and delimits the turbulence strength above which the scintillation index exceeds unity \cite{Andrews2005}.

To test our model, a Monte Carlo simulation was performed using randomly generated phase screens, along the lines of Ref. \cite{Schmidt2010}, and using equations \eqref{ampli2} and \eqref{ampli3} to describe propagation of the coincidence beams. The resulting curves are shown in Fig. \ref{fig:results}. Again, Kolmogorov spectrum was assumed, along with turbulence strength homogeneity inside the chamber, leaving only the number of phase screens and their separation as free parameters; we chose these to be the same for all curves. While there are some discrepancies, which we tentatively attribute to differences between the refractive index statistics in our turbulence chamber and that of the atmosphere, the normalized power increase in the inverted coincidence beam scenario closely matches what was measured.

Additionally, curve fits (not shown) were performed on the data by considering the Gaussian width to broaden according to $w_{LT}=w_0\left[1+\alpha (k^{1/3}/w_0^{5/3})\gamma^2\right]^{1/2}$ with $k$ and $w_0$ corresponding to each data set. The fit parameter $\alpha$ can be interpreted as the effective turbulence response and provides us with a figure of merit. The total signal going through the aperture is then $S_L(\gamma) = \mathrm{erf}\left( \frac{a}{\sqrt{2} w_{LT}}\right)$ for the laser measurements, $a=50\ \mu$m being the slit width, and $S_C(\gamma) = 1/w_{LT} \int_{-\infty}^{\infty} \exp\left(-2x^2/w^2_{LT}\right) \Lambda(2x/a)dx$ for the coincidence measurements, where $\Lambda(x)=1-|x|$ for $|x|<1$ and $0$ otherwise \cite{Walborn2010}. The fitted curves are therefore $S_L(\gamma)/S_L(0)$ and $S_C(\gamma)/S_C(0)$.

\begin{figure}[t]
\centering
\includegraphics{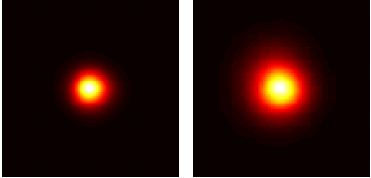}
\caption{Averaged He-Cd beam profile with and without tilt correction.\label{fig:HeCd}}
\end{figure}
\begin{table}[ht]
\centering
\begin{tabular}{c|c}
\hline\hline
Setup & $\alpha$ \\
\hline\hline
Coincidences with inversion   & 0.10 $\pm$ 0.03\\
Coincidences without inversion & 1.01 $\pm$ 0.11\\
He-Cd without correction  & 1.02 $\pm$ 0.08\\
He-Ne without correction  & 1.07 $\pm$ 0.09 \\
Corrected He-Cd  & 0.21  $\pm$ 0.02\\
Corrected He-Ne & 0.30  $\pm$ 0.04\\
\hline
\end{tabular}
\caption{Fit parameters $\alpha$ corresponding to the different setups (see text).}
\end{table}

An additional test was performed by correcting both lasers for wavefront tilt, via post-processing. This is done by displacing the beam spot transversely in each frame of the recorded video so that its centroid always falls on the same position, before taking the average over the frames. The pictures in Fig. \ref{fig:HeCd} show the improvement with this correction applied to the He-Cd beam for $\gamma = 27.9 \ \mu$m (maximum turbulence strength). All fit parameters $\alpha$, including the ones corresponding to tilt-corrected data, are shown on Table~II.

The ratios between $\alpha$ for the corrected and uncorrected lasers ($0.21 \pm 0.03$ and $0.28 \pm 0.04$ for He-Cd and He-Ne) agree reasonably well with the value of $0.25$ from models found in the literature \cite{Andrews2005}. It is to be noticed that although a significant improvement can be made just by correcting wavefront tilt ($80\%$ for the He-Cd and $71\%$ for the He-Ne), it is not as significant as the improvement reached with the inversion on the coincidence beam ($90\%$), thus providing further support to the claim that aberrations other than tilt are being corrected.

\section{Conclusion}

An interesting question that naturally arises is whether entanglement is actually necessary to attain the cancellation effect observed. In Ref. \cite{Bennink2002} it is demonstrated that for a known transfer function, classical correlations can be engineered in such a way as to reproduce any joint detection probability attainable in a single plane with entangled photons. Since the channel transfer function changes in an unpredictable way because of turbulence, it would need to be continuously monitored in order to have its form determined, in the same lines of adaptive optics. In our case, the necessary conditions are the transfer of angular spectrum from the pump beam to the down-converted two-photon field, which allows for the control of the spatial correlations in the far field, associated with a strong ($\delta$-like) spatial correlation of the two photons on the source plane. Control of correlations in both far and near fields is naturally limited with separable two-photon sources \cite{Bennink2004}. Moreover, in order to take practical advantage of the effect reported here, the single-count rate or intensity transverse profile observed on the detection plane must be much broader than the coincidence rate profile. Since the ratio between the single and coincidence count profile widths gives a a good estimate for the Schmidt number of the two-photon state entangled in spatial modes \cite{Fedorov2007}, our scheme would work well only for highly entangled states. In this sense, our results suggest that spatial mode entanglement can be protected against turbulence when two-photon states are transmitted through the atmosphere.

\section*{Acknowledgements}
We acknowledge the support from the brazilian funding agencies CNPq, CAPES and FAPEMIG.

\end{document}